\documentclass{webofc}
\usepackage[varg]{txfonts}   
\usepackage{listings}
\usepackage{hyperref}
\newcommand{\Ks}{{K_{\rm s}}}
\newcommand{\JHK}{{JHK_{\rm s}}}
\newcommand{\AK}{{A_{K_{\rm s}}}}
\newcommand{\EHK}{{E_{H-K_{\rm s}}}}

\newcommand{\AKEHK}{{A_{K_{\rm s}}/E_{H-K_{\rm s}}}}
\newcommand{\AKEJK}{{A_{K_{\rm s}}/E_{J-K_{\rm s}}}}
\begin{document}
%
\title{Time-Series Surveys and Pulsating Stars:\\
The Near-Infrared Perspective}
%
%

\author{\firstname{Noriyuki} \lastname{Matsunaga}\inst{1}}

\institute{Department of Astronomy, The University of Tokyo, 7-3-1 Hongo, Bunkyo-ku, Tokyo 113-0033, Japan}

\abstract{%
The purpose of this review is to discuss the advantages and problems
of near-infrared surveys in observing pulsating stars in the Milky Way.
One of the advantages of near-infrared surveys, when compared to
optical counterparts, is that the interstellar extinction is
significantly smaller. As we see in this review, a significant volume of
the Galactic disk can be reached by infrared surveys but
not by optical ones.
Towards highly obscured regions in the Galactic mid-plane, however,
the interstellar extinction causes serious problems even with near-infrared data
in understanding the observational results.
After a review on previous and current near-infrared surveys,
we discuss the effects of the interstellar extinction in 
optical (including {\it Gaia}) to near-infrared broad bands based
on a simple calculation using synthetic spectral energy distribution.
We then review the recent results on classical Cepheids
towards the Galactic center and the bulge, as a case study,
to see the impact of the uncertainty in the extinction law.
The extinction law, i.e.~the wavelength dependency of the extinction,
is not fully characterized, and its uncertainty makes it hard
to make the correction. Its characterization is an urgent task
in order to exploit the outcomes of ongoing large-scale surveys of pulsating
stars, e.g.\  for drawing a map of pulsating stars across the Galactic disk.
}
\maketitle

\section{Introduction}\label{sec:intro}

What we call ``near-infrared'' (hereafter near-IR) here 
is the wavelength range covered by the photometric bands of $JHK$.
A broad and general review on the photometric bands is found in
\cite{Bessell-2005}. Note that the $K$ band, at around 2~$\mu$m,
has variations of filter transmission in different systems, and  
we mainly consider the $\Ks$ band which lacks the longest-$\lambda$ part
($\lambda > 2.3~{\mu}$m) of the broader $K$ in the following discussions.
The wavelength around 3.5~{$\mu$}m will be called ``mid-IR'',
partly because most important datasets in this range tend to
be collected with space facilities rather than ground-based telescopes today,
but we also discuss some results obtained in this mid-IR range.
A review on studies in the shorter-$\lambda$ ($I$-band and shorter) range
is given by Soszy\'nski in this volume (\cite{Soszynski-2017}).

There are mainly three advantages of IR data
for observations of pulsating stars.
First, they are less affected by interstellar extinction.
At around 2~{$\mu$m}, for example, the extinction is
around 10~\% of that at the optical $V$ band
(\cite{Rieke-1985,Cardelli-1989}) or even smaller
({$\sim$}1/16, \cite{Nishiyama-2008}).
Secondly, some objects are enshrouded in circumstellar dust,
so that they are only visible or bright in the IR
(\cite{Tanabe-1997,Tanabe-2004}).
IR data are crucial to study properties of such dust.
Thirdly, we know pulsating stars tend to show more simple characteristics
in the IR. In particular, the period-luminosity relation of Cepheids
are known to be tight and less affected by metallicity.
There has been a wealth of literature on its advantage
(e.g.~\cite{Madore-1991,Bono-2010,Inno-2013,Scowcroft-2016,Bhardwaj-2016a}).
This is of great value for high precision cosmology (\cite{Riess-2016})
and for other applications.

The following five facilities have been actively used for
observing pulsating stars in the IR: Infrared Survey Facility (IRSF)
in South Africa, VISTA in Chile,
{\it Hubble Space Telescope}, {\it Spitzer Space Telescope},
and CTIO 1.5-m telescope with CPAPIR camera. Catalogs from all-sky surveys
like 2MASS\footnote{http://irsa.ipac.caltech.edu/Missions/2mass.html}
and WISE\footnote{http://irsa.ipac.caltech.edu/Missions/wise.html}
are very conveniently used and naturally making a giant impact
on studies of pulsating stars, but they are not discussed in this review.

\begin{itemize}

\item
The 1.4-m telescope IRSF is working with SIRIUS near-IR camera
in the SAAO Sutherland observatory, South Africa
(\cite{Nagashima-1999,Nagayama-2003}).
Even with the moderate field-of-view, $7.7^{\prime}\times 7.7^{\prime}$,
the efficiency of the three detectors taking simultaneous images
in three bands ($\JHK$) has been producing important catalogs
of variable stars with color information for many stellar systems and regions:
the Galactic center and bulge
(\cite{Matsunaga-2009,Matsunaga-2011,Matsunaga-2013,Matsunaga-2016}), 
globular clusters (\cite{Matsunaga-2006,Sloan-2010}),
the Magellanic Clouds (\cite{Ita-2004a,Ita-2004b}),
and nearby dwarf galaxies (\cite{Menzies-2002,Menzies-2008,Whitelock-2009,Menzies-2010,Menzies-2011,Feast-2012,Whitelock-2013,Menzies-2015}).
Some of these results will be discussed in more detail below.

\item
VISTA is a 4.1-m telescope in the ESO Cerro Paranal observatory, Chile, and
is working with a wide-field near-IR camera VIRCAM.
The sixteen detectors, 67 million pixels effectively covering
${\sim}0.6$~deg$^2$, together with the relatively large telescope aperture
have a very high survey ability (\cite{Sutherland-2015}).
Among the six public surveys, two surveys are targeted at surveying
pulsating stars: VISTA Variables in the Via Lactea (VVV, \cite{Minniti-2010})
and VISTA survey of the Magallanic Clouds (VMC, \cite{Cioni-2011}).
See also Minniti {et~al.} \cite{Minniti-2017c} and Cioni {et~al.}
\cite{Cioni-2017} in this proceedings.
While the comprehensive photometric catalog provided by the VVV
has been used in many investigations about the Galactic bulge and disk
(\cite{Saito-2011,Minniti-2011,Gonzalez-2012,Gonzalez-Fernandez-2012,Borissova-2014,Valenti-2016}),
its time-series data have led to new discoveries and
insights into various pulsating stars like 
classical Cepheids (\cite{Dekany-2015a,Dekany-2015b,Chen-2017}) and
RR Lyrs (\cite{Dekany-2013,Alonso-Garcia-2015,Gran-2015,Gran-2016,Minniti-2016,Minniti-2017a,Minniti-2017b}).
VMC has been also providing very useful near-IR datasets
for pulsating stars, 
classical Cepheids (\cite{Ripepi-2012,Moretti-2016,Ripepi-2016,Marconi-2017}),
type II Cepheids (\cite{Ripepi-2015}), and RR Lyrs (\cite{Moretti-2014})
as well as other topics such as star formation history 
in the Magellanic Clouds (\cite{Kerber-2009,Rubele-2012,Rubele-2015}).
In addition to time variations in brightness, 
both of the VVV and VMC surveys are providing proper motions
\cite{Cioni-2014,Cioni-2016,Gromadzki-2016,Kurtev-2017}
which will be very useful to discuss the nature of variable stars
and other populations.
Besides the two large surveys, there is an important contribution
(and will be more) from VISTA:
McDonald {et~al.} \cite{McDonald-2013,McDonald-2014}
studied variable stars in the Sagittarius dwarf spheroidal galaxy.

\item
{\it Spitzer Space Telescope} is actively used
to obtain important time-series datasets in the mid IR
through some conspicuous projects.
The Carnegie Chicago Hubble Program has been making a considerable effort
on establishing the period-luminosity relation of Cepheids in the mid IR 
(see e.g.~\cite{Freedman-2011,Freedman-2012,Scowcroft-2013,Scowcroft-2016}).
The {\it Spitzer} Legacy Program ``Surveying the Agents of
a Galaxy's Evolution'' (SAGE, \cite{Meixner-2006}) and its related programs
(SAGE-Var \cite{Riebel-2015}, in particular) collected a comprehensive
dataset in the mid IR for pulsating stars and other objects in
the Magellanic Clouds \cite{Riebel-2010,Riebel-2015,Polsdofer-2015}.
In this proceedings book, Whitelock {et~al.} \cite{Whitelock-2017} discuss
mid-IR characteristics
of large amplitude variables in the LMC and IC~1613 by combining
the SAGE-Var data and other datasets. 

\item
{\it Hubble Space Telescope (HST)} has been also producing important results
on pulsating stars mainly for the purpose of cosmology and
determining the $H_0$ constant in particular.
While the {\it HST} Key Project to measure $H_0$ in the 1990s used
optical data for detecting Cepheids in distant galaxies as much as possible
(\cite{Freedman-2001} and references therein), Riess and collaborators
have been using near-IR data exclusively collected with
an identical instrument to reduce the systematic uncertainties
caused by the intrinsic characteristics of Cepheids like
the metallicity effect on the period-luminosity relation and
by the cross-instrument errors (\cite{Riess-2016}).

\item
The CPAPIR near-IR camera attached to the CTIO 1.5-m telescope
was used to carry out the LMC Near-IR Synoptic Survey (\cite{Macri-2015}).
Its good collection of time-series data for a wide area of the LMC
has been producing important results on classical Cepheids
and type II Cepheids
(\cite{Bhardwaj-2015,Bhardwaj-2016a,Bhardwaj-2016b,Bhardwaj-2017}).

\end{itemize}

It is clear that these surveys are providing us with useful data
on pulsating stars. The IR surveys are particularly useful
for exploring the large space of the Milky Way including
the disk obscured by interstellar extinction, which is the main focus
of the following discussions.
When interesting and reddened objects are found in obscured regions,
follow-up spectroscopy in the IR will be important. The demands for
IR spectrographs will grow rapidly considering that a large number
of pulsating stars and other objects are being found in IR surveys.
APOGEE and its successor APOGEE-2 (\cite{Majewski-2016})
together with other modern near-IR high-resolution spectrographs
like GIANO \cite{Origlia-2014} and WINERED \cite{Ikeda-2016} will play
important roles in collecting detailed information such as
radial velocities and chemical abundances of
pulsating stars, but this is beyond the scope of this review.

\section{Expected limits of optical and near-IR surveys}

In this section, we discuss limits of optical and near-IR surveys
by calculating broad-band photometry with different extinction laws.
We consider classical Cepheids and examine how far surveys can reach.
Previous surveys of Cepheids are far from complete (see, e.g.\  figure~1
in \cite{Matsunaga-2012}), which is mainly due to the interstellar extinction.
Windmark {et~al.} \cite{Windmark-2011} predicted the total number 
and the distribution of Cepheids which can be detected in
the {\it Gaia} survey. They obtained {$\sim$}20000 
based on a simple exponential-disk
model\footnote{Such a simple exponential disk,
see their prediction in figure~5 of \cite{Windmark-2011},
is not consistent with the distribution of Cepheids we found
as described in section \ref{sec:discussion}.}
and the local density.
They also predicted about half of them will be detected by {\it Gaia}
with the limiting magnitude of $G=20$~mag.
While {\it Gaia} will detect a large number of new Cepheids,
such an optical survey is affected by
the interstellar extinction and is limited up to several kpc along
the Galactic plane. The accurate limit, however, depends on 
the extinction law as we see below.

\begin{figure}
\centering
\includegraphics[width=9cm,clip]{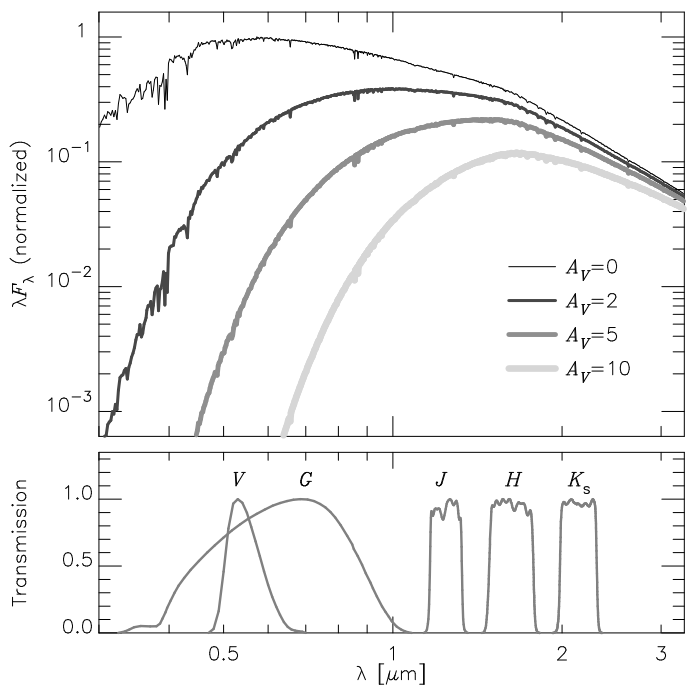}
\caption{(Top)~The synthetic spectrum of the Sun based on ATLAS9 by Kurucz and Castelli and those affected by the power law extinction, $\sim \lambda^{-2}$. (Bottom)~Filter transmission curves, normalized to the peak of each bandpass.}
\label{fig:SED}       
\end{figure}

In order to see the effects of the extinction on the broad-band photometry,
we present results of integrating the spectral energy distribution
after the extinction applied. This calculation is summarized as
\begin{equation}
m_\lambda = \int F_\lambda A_\lambda T_\lambda d\lambda,
\end{equation}
where $F_\lambda, A_\lambda$, and $T_\lambda$ indicate
flux density, wavelength-dependent extinction function, and
filter transmission, respectively. 
For $F_\lambda$, we adopt the synthetic spectrum\footnote{
The solar spectrum obtained by using the ATLAS9 code with the solar abundance
in \cite{Asplund-2005} was taken from http://wwwuser.oats.inaf.it/castelli/sun.html}
of the Sun whose effective temperature is within
the range of Cepheids' temperature. For comparing the effects of
different extinction laws, we here consider the power law,
$A_\lambda \propto \lambda^{-\alpha}$, with
$\alpha=2$ and the law of Cardelli \cite{Cardelli-1989}.
The former is close to the extinction law in
Nishiyama {et~al.} (2006, \cite{Nishiyama-2006}),
which we call the N06 law hereinafter, while
the Cardelli law (henceforth C89) corresponds to the power law
with $\alpha=1.61$,
significantly shallower than the N06 law.
We consider the filters, $V, G, J, H$ and $\Ks$ for $T_\lambda$.
The transmission curves are taken from \cite{Bessell-2005} for $V$,
\cite{Jordi-2010} for $G$, and the technical information on the SIRIUS camera
for $\JHK$.
Figure~\ref{fig:SED} plots the SED
with varying amount of extinction, i.e.\  $F_\lambda A_\lambda$,
from $\AK=0$ to 10~mag in addition to the filter transmission,
$T_\lambda$.
Note that in our calculation the integration of the SED is done after
the $\lambda$-dependent extinction is applied.

\begin{figure}
\centering
\includegraphics[width=9cm,clip]{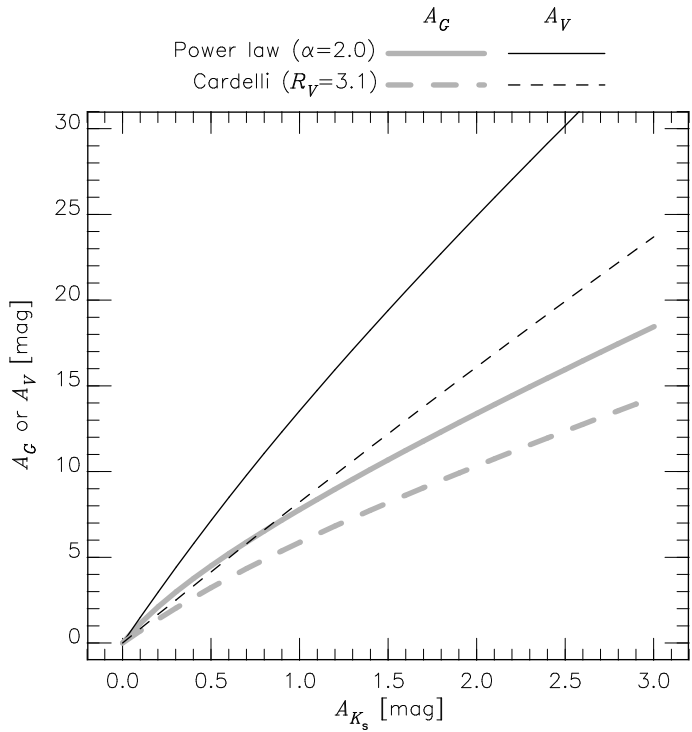}
\caption{The optical extinctions, $A_G$ (gray) and $A_V$ (black),
are plotted against $\AK$. The filled curves are obtained for
the power law with $\alpha=2$ while the dashed curves for the Cardelli law
with $R_V=3.1$ as indicated in the legend.
}
\label{fig:ext_GVK}       
\end{figure}

Figure~\ref{fig:ext_GVK} plots the extinctions in $V$ and
{\it Gaia}'s $G$ bands against that in $\Ks$. 
For example, $A_V/\AK$ is around 14 with the N06 law (almost consistent
with the result in \cite{Nishiyama-2008}, {$\sim$}16) and around 8 with
the C89 law. The extinction in $G$ is clearly smaller than that in $V$.
{\it Gaia} can reach relatively deeper than previous $V$-band surveys
in obscured regions of the Milky Way. Furthermore, the curves for
$A_G/\AK$ in Figure~\ref{fig:ext_GVK} gets shallower towards
the larger $\AK$. This is because the $G$ band is broad and
the effective wavelength changes with increasing the amount of extinction.
The central wavelengths of $G$ and $V$ bands are
0.673 and 0.551~$\mu$m respectively \cite{Jordi-2010}.
The SED affected by a large extinction, however, keeps more photons
in the longer wavelength part of the $G$ band,
as illustrated in Figure~\ref{fig:SED}, 
so that the sensitivity to the extinction gets closer to that of
a photometric band at for a longer-$\lambda$ range.
This leads to an even larger difference between
$A_V/\AK$ and $A_G/\AK$ at a larger $\AK$; at around $\AK=2.5$~mag,
$A_G$ gets nearly half of the corresponding $A_V$.

\subsection{How deep will {\it Gaia} and optical surveys be able to reach?}
\begin{figure}
\centering
\includegraphics[width=9cm,clip]{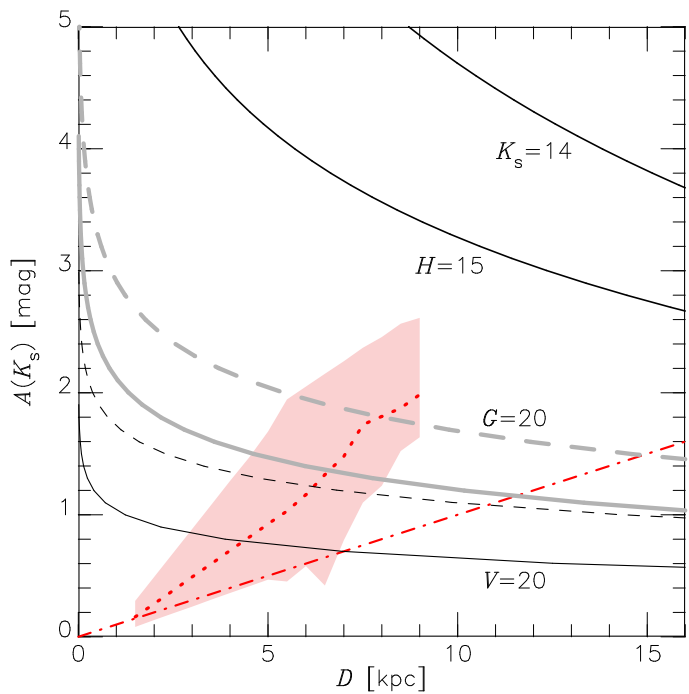}
\caption{
The maximum extinction, $\AK$, with which a Cepheid with $P=10$~days
can be detected with a given limiting magnitude.
The limiting magnitudes are given within
the figure, while a Cepheid with $P=10$~days has
absolute magnitudes of $-4, -4, -5.6$, and $-5.7$~mag in $VGH\Ks$ bands
respectively. For $V$ and $G$, the same styles are used as in
Figure~\ref{fig:ext_GVK}: filled curves when the power law with $\alpha=2$
is used and dashed curves with the Cardelli law used.
Only the power law is used for the $H$ and $\Ks$ bands.
The red dotted curve and the shaded area illustrate
the three-dimensional extinction map by \cite{Schultheis-2014} towards the Bulge
(see also Figure~3 of \cite{Matsunaga-2016}), while dashed-dotted line
indicates the extinction increasing at the rate of 0.1~mag/kpc.
}
\label{fig:DAKrange}      
\end{figure}

Figure~\ref{fig:DAKrange}
plots the maximum extinction that can be reached with a given magnitude
in each photometric band as a function of distance.
The limiting magnitudes of $G=20$,
$V=20, H=15$, and $\Ks=14$~mag are adopted as an example.
The interstellar extinction varies from one line of sight to another,
and we here consider two cases: the three-dimensional extinction map
towards the bulge taken from \cite{Schultheis-2014} and
a moderate increase of extinction at the rate of 0.1~mag/kpc.
With the N06 law adopted (indicated by the gray filled curve),
the {\it Gaia}'s limiting magnitude (\cite{vanLeeuwen-2017}) allows
one to detect Cepheids with $\AK\sim 1.2$~mag at 
${\sim}$6.5~kpc towards the bulge or at over 10~kpc in the latter case
of the moderate extinction.
These limits are significantly deeper than a $V$-band survey.
If the extinction law were closer to the C89 law, the limits would be
even deeper, but at least towards the bulge
the N06 law is considered to be more likely (\cite{Nishiyama-2008}). 
Towards less obscured regions, {\it Gaia} and some other optical surveys
can see to larger distances. 
The sensitivity of {\it Gaia} and more importantly
its all-sky coverage will extend the frontier of the variability survey.
Other current and future optical surveys will also make important
contributions to finding new variables and exploring their population
in the wide range of the Milky Way.
For example, Feast {et~al.} \cite{Feast-2014} identified classical Cepheids,
among OGLE-III Cepheids (\cite{Soszynski-2011}), which are located
in the flared part of the Galactic disk beyond the bulge. 
These objects are slightly off the Galactic mid-plane ($2<|b|<5$~deg).
The extinctions of these objects, 0.17--0.57~mag, at the distances
of 22--30~kpc correspond to roughly 0.01~mag/kpc but with
a scatter of 50~\%. 
The OGLE-IV survey for the Galactic disk
has actually revealed a rich population of Cepheids
in the southern hemisphere as presented by Udalski {et~al.}
in this volume \cite{Udalski-2017}. In their maps, however, newly found
Cepheids and RR Lyrs have patchy distributions which clearly
shows the limit of optical surveys and the needs of IR surveys.

\subsection{How deep can near-IR surveys reach?}
 Even with the moderate limiting magnitudes,
$H$- and $\Ks$-band surveys can reach significantly further than 10~kpc
except through thick molecular clouds. The extinctions of objects towards
the Galactic center are {$\sim$}2.5~mag in $\AK$, which makes
it impossible to detect them in the optical ranges.
The {\it Gaia} all-sky map, accompanying its first data release,
clearly shows the dark lane caused by the disk extinction
(figure~2 in \cite{Gaia-2016}).
The Galactic plane within 1~deg tends to have the extinctions
around 1~mag in $\AK$ or stronger \cite{Dutra-2003,Schultheis-2014}
and thus near-IR surveys are
more effective than optical surveys at around 5~kpc and further.
As mentioned in the Introduction, 
some results from near-IR surveys carried out with IRSF and VISTA
have already demonstrated that many obscured variables can be
found in a large space of the Galactic mid-plane.
Such surveys seeing through the Galactic plane is necessary,
in particular, to study the thin disk component;
the Galactic latitude of 1~deg corresponds to
the distance of 140~pc from the plane at the distance of 8~kpc.
It is difficult to predict how much we can ultimately
see through the entire range of the disk. 
Towards the Galactic center and bulge, for example,
if the same amount of extinction needs to be overcome
on the opposite side of the disk,
we would expect the extinction of $\AK\sim 5$~mag.
It is possible to observe such reddened objects
if they are intrinsically bright, although thick molecular clouds may well
interrupt the lines of sight to prevent us from reaching to the opposite end
even in the IR.

\section{Impact of the uncertainty in the extinction law}\label{sec:discussion}

\begin{figure}
\centering
\includegraphics[width=11cm,clip]{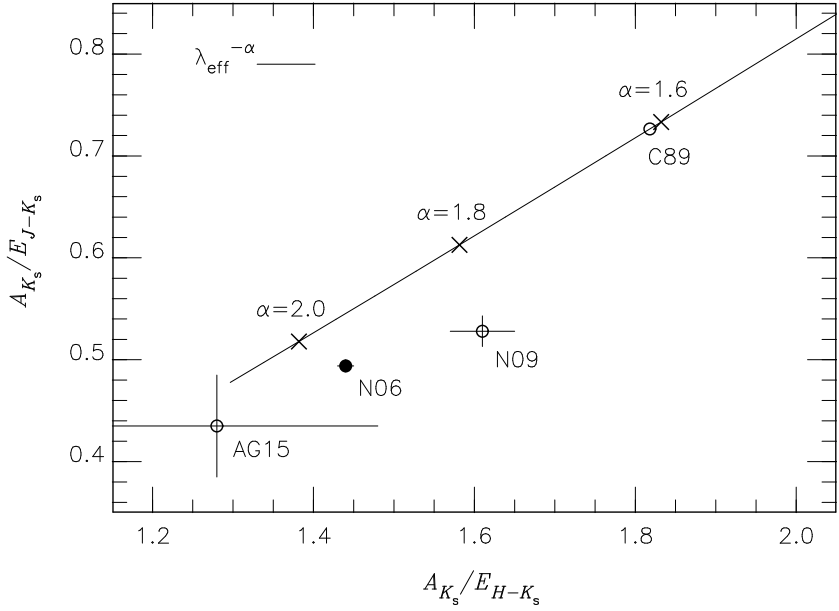}
\caption{Comparison of the extinction coefficients, $\AKEJK$ and $\AKEHK$.
The straight line gives the relation between the two coefficients
with varying power $\alpha$ as indicated by crosses.
The points with the labels of AG15, C89, N06, and N09 indicate
the values obtained by \cite{Alonso-Garcia-2015}, \cite{Cardelli-1989},
\cite{Nishiyama-2006}, and \cite{Nishiyama-2009}, respectively.
The C89 law corresponds to the power law with $\alpha=1.61$.
}
\label{fig:AEAE}       
\end{figure}

\begin{figure}
\centering
\includegraphics[width=13.5cm,clip]{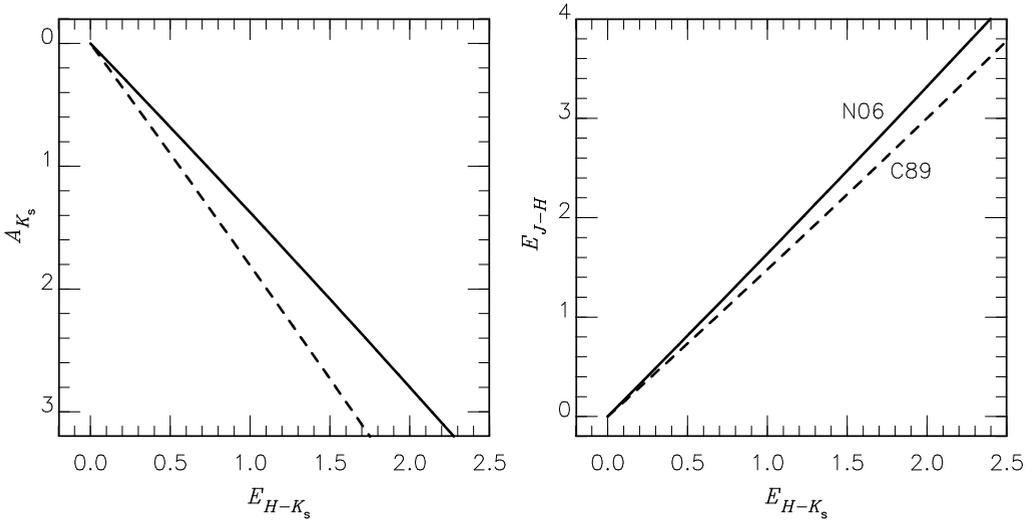}
\caption{Reddening vectors according to the extinction laws of
C89 (\cite{Cardelli-1989}, dashed) and N06 (\cite{Nishiyama-2006}, solid)
on the color-magnitude diagram (left) and the color-color diagram (right).}
\label{fig:ext_JHK}       
\end{figure}

The near-IR surveys will open our path to the obscured regions of the Milky Way
as we discussed in the previous section, but the interstellar extinction
still poses a serious problem with interpreting observational data
we'll obtain. For many applications, we need to make corrections of
the extinction and reddening, and this requires using a proper law
of extinction. When we combine $H$- and $\Ks$-band data with
the period-luminosity relation of Cepheids
to estimate the distances and foreground extinctions, for example,
we need to derive the amount of extinction, $\AK$, considering
an extinction coefficient, $\AKEHK$, and the reddening,
$\EHK=(H-\Ks)-(M_H-M_{\Ks})$, where $M_H$ and $M_{\Ks}$ are predicted 
by the period-luminosity relation\footnote{When one uses the relation of a Wesenheit index, e.g.~$W_{\Ks}=\Ks-\alpha (H-\Ks)$, which is often considered as a reddening-free index, it is still necessary to decide the coefficient $\alpha$ properly in order to make this index reddening free.}. 
Figure~\ref{fig:AEAE} plots some extinction coefficients
in the near IR in literature and shows a large scatter among them.
Figure~\ref{fig:ext_JHK} illustrates how the different extinction laws
(N06 and C89) give different reddening vectors on the color-magnitude diagram
(left) and the color-color diagram (right). 
For example, the reddening of $\EHK=1.5$~mag would give
$\AK=2.16$ and 2.73~mag with the N06 and C89 laws, respectively.
The differences in the extinction laws thus introduce large uncertainties
in estimating distances of such reddened objects (the reddenings of
stars at around the Galactic center are around 1.5--2~mag in $\EHK$).
On the other hand, Figure~\ref{fig:ext_JHK} also indicates that
the reddening vector on the color-color diagram is relatively insensitive to
the difference in the extinction law; the N06 and C89 laws would predict
different $(H-\Ks,J-H)$ by only ${\sim}$0.2~mag for the stars at around
the Galactic center.

A good example of the impact of the extinction law is found
and discussed in \cite{Matsunaga-2016}.
D\'ek\'any {et~al.} \cite{Dekany-2015b} reported a few dozens of
classical Cepheids towards the Galactic bulge and concluded that
they form a thin disk surrounding the Galactic center.
Matsunaga {et~al.} \cite{Matsunaga-2016}, however, reported their
discoveries of classical Cepheids which are located behind the bulge
and concluded that the region around the Galactic center lacks
classical Cepheids.
Some of the Cepheids are common in the two works, but they obtained
significantly different distances for the Cepheids, $\Delta\mu_0 =0.4$--0.5~mag.
Matsunaga {et~al.} \cite{Matsunaga-2016} found that such a large difference
was caused because two different extinction laws were used:
the N06 law in \cite{Matsunaga-2016} and the N09 law from
Nishiyama {et~al.} (2009, \cite{Nishiyama-2009}) in \cite{Dekany-2015b}. 
The $\AKEHK$ in the N06 and N09 laws are 1.44 and 1.61 (Figure~\ref{fig:AEAE}),
respectively, and these led to offsets of {$\sim$}0.3~mag and
the significantly different conclusions on the distribution of the Cepheids.

\begin{figure}
\centering
\includegraphics[width=13.5cm,clip]{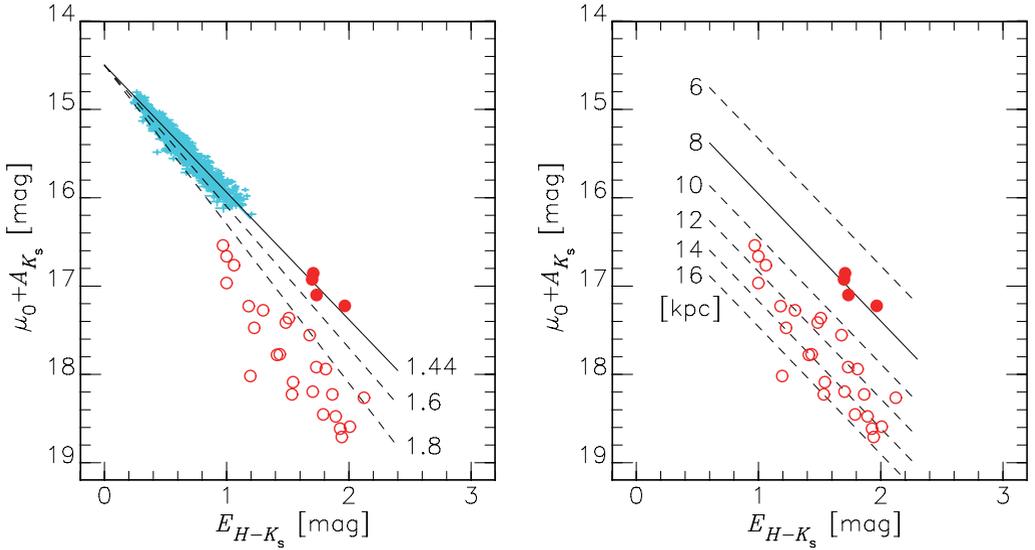}
\caption{(Left) Reddening vectors on the color-magnitude diagram
with different coefficients of $\AKEHK$
starting from the distance of the Galactic center, 8~kpc, at
the zero color excess, $\EHK=0$~mag. The sequence of cyan points
near the top-left corner, taken from N06 \cite{Nishiyama-2006},
indicates the red clump peaks affected by various amounts of reddening. 
Filled circles indicate
the four classical Cepheids in the nuclear stellar disk
(\cite{Matsunaga-2011,Matsunaga-2015}), and open circles indicate
the other classical Cepheids reported in \cite{Matsunaga-2016}.
(Right)~Same as the left panel, but lines have the slope of
$\AKEHK=1.44$ and correspond to different distances from 6 to 16~kpc.}
\label{fig:CMDsCC}       
\end{figure}

Readers are referred to \cite{Matsunaga-2016} for details, but we here
outline how the extinction law towards the Galactic bulge can be
constrained together
with adding a new plot which supports our conclusion.
An important group of Cepheids are four classical Cepheids which
belong to the nuclear stellar disk (NSD, henceforth).
The NSD is a disk-like system around the Galactic center, 
with a radius of about 200~pc. This region is also known as
the Central Molecular Zone (CMZ). Unlike the more extended bulge, this region
is known to host young stars like the famous massive clusters Arches
and Quintuplet, and those young stars are rotating within this disk. 
We can conclude that the four Cepheids belong to the NSD for a few reasons:
the projected distances, similarity in the physical parameters of Cepheids,
the fact that their distances must be similar independent of
the adopted extinction law, and that their radial velocities
are consistent with the rotation of the NSD.
These strongly indicate that their distances should agree with
that of the Galactic center ($R_0 =8.0\pm 0.5$~kpc,
\cite{Gillessen-2013,deGrijs-2016}) within the size of
the NSD (${\sim}0.2$~kpc). Note that some estimates of $R_0$,
e.g.\  the orbit modeling of the S stars around the central blackhole
\cite{Gillessen-2013}, are independent of the extinction law.
As illustrated in the left panel of Figure~\ref{fig:CMDsCC},
$\AKEHK$ should be around 1.44, rather than 1.61, in order to keep
the four Cepheids at around 8~kpc.
Note that, with the period-luminosity relations and observed magnitudes given,
the location of each Cepheid on this diagram doesn't depend on
the adopted extinction law or its distances.
The four NSD Cepheids are precisely on the extension of the Nishiyama's fit
to the red clumps. Assuming that the Cepheids and the red clumps are
at the same distance, this strongly supports the extinction value of 1.44.
Once the coefficient of 1.44 is adopted,
the distances of other Cepheids can be determined as one can see
on the right panel of Figure~\ref{fig:CMDsCC}. 
The vertical offset in this panel corresponds to the offset in distance.
Other Cepheids are located significantly further than the Galactic Center,
showing the gap between the NSD (marked down by the four Cepheids) and
the inner edge of the distribution of disk Cepheids.
Also, we found no Cepheids on the nearer side, which supports
the absence of Cepheids around the Galactic Center;
the saturation limit of our survey could have allowed us to
detect Cepheids between ${\sim}$5 and 8~kpc from the Sun if
the density were high enough. These support our conclusion on
the lack of Cepheids within 2.5~kpc around 
the Galactic center except the NSD. 
More details are given in the original paper \cite{Matsunaga-2009}
as mentioned above,
and some remarks are also found in the conference summary 
for this conference \cite{Feast-2017}.

\begin{acknowledgement} 
\noindent\vskip 0.2cm
\noindent {\em Acknowledgments}:
The author is grateful to Michael Feast for reading the draft and for
his useful comments. We appreciate financial support from
the Japan Society for the Promotion of Science (JSPS) through
the Grant-in-Aid, No.~26287028.

\end{acknowledgement}

%
%

\end{document}